\documentclass[%
 reprint,
 amsmath,amssymb,
 aps,
 prl,
floatfix,
]{revtex4-1}

\usepackage{igo}
\usepackage{graphicx}% Include figure files
\usepackage{setspace}
\usepackage{dcolumn}% Align table columns on decimal point
\usepackage{minibox}
\usepackage{bm}% bold math
\usepackage{hyperref}% add hypertext capabilities
\usepackage{float}%positions the figures
\floatstyle{boxed} 
\usepackage{natbib}
\usepackage{url}

\newcommand{\bra}[1]{\ensuremath{\left\langle#1\right|}}
\newcommand{\ket}[1]{\ensuremath{\left|#1\right\rangle}}

\begin{document}

\title{Quantum Go}% Force line breaks with \\

\author{Andr$\acute{e}$ Ranchin$^{1,2}$ }

\affiliation{$^{1}$University of Oxford, Department of Computer Science, Quantum Group}
\affiliation{$^{2}$Imperial College London, Department of Physics, Controlled Quantum Dynamics}

\date{\today}% It is always \today, today,

\begin{abstract}

We introduce a new board game based on the ancient Chinese game of \textbf{Go} (Weiqi, Igo, Baduk). The key difference from the original game is that players no longer alternatively play single stones on the board but instead they take turns placing \textbf{pairs of entangled go stones}. A phenomenon of \textbf{quantum-like collapse} occurs when one of the stones in an entangled pair is directly in contact with at least one other stone. The player to whom the entangled pair belongs must then remove one of the stones in the pair from the board. The aim of the game is still to surround more territory than the opponent and as the number of stones increases, all the entangled pairs of stones eventually reduce to single stones. \textbf{Quantum Go} provides an interesting and tangible illustration of quantum concepts such as \textit{superposition}, \textit{entanglement} and \textit{collapse}.

\end{abstract}

\maketitle

\section{Introduction}

In light of the recent defeat of Lee Sedol by the Google DeepMind computer program AlphaGo \cite{guru} and the incredible success of neural network methods \cite{Sil16} in producing a top professional-level go program, it is interesting to study variants of the game which may be even more challenging to master for a computer player. A natural extension of the game of Go is to play on boards whose size is larger than the traditional 19 $\times$ 19 grid or that have alternative shapes and topological properties \cite{Mil,Har}.

In this article, we will introduce a novel variant of the game of Go, called \textbf{Quantum Go}. This is similar in spirit to the games of Quantum Tic-Tac-Toe \cite{All06} and Quantum Chess \cite{Sel10, Chess} and provides an entertaining and compelling platform to introduce fundamental quantum concepts such as \textit{entanglement}, \textit{superposition} and \textit{collapse}.

The basic rules of the game are very close to those of the original game of Go except for a key difference: players no longer alternatively place single stones on the board. Instead, both players take turns placing pairs of stones on the board, which we call entangled pairs. Both stones in an \textbf{entangled pair} remain on the board until there is a stone on one of the intersections directly touching one of the stones in the pair. At that point, all the entangled pairs containing a stone adjacent to either of the stones just played, as well as the last entangled pair itself, will \textbf{collapse}, in the sense that both players must remove one stone from each entangled pair. The game then proceeds as before. \\

The following example, shown in Figure 1, illustrates how entangled pairs of stones collapse to single stones. Note that we use the following coordinate notation throughout the article: columns are labeled from A (left) to T (right), skipping I, and rows are labeled from 1 (bottom) to 19 (top). Black first places a pair of entangled stones \blackstone[1] at $\{D3, C5\}$ then white places a pair of entangled stones \whitestone[2] at $\{D2, E6\}$. 

\black[1]{c5} \white[2]{d2}   
\black[1]{d3} \white[2]{e6} 
\gobansymbol{c2,d1,e2}{x}
\begin{center}
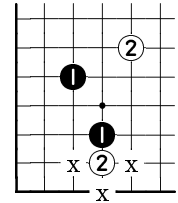
\shortstack{\showgoban\\Figure 1: Before the collapse of two pairs of entangled stones}
\end{center}

Given that the four neighboring intersections of the D2 white stone contain the D3 black stone (as well as the three intersections marked with an $\times$), both pairs of entangled stones collapse.

The white player has no stones touching the entangled pair that she has just played so the black player first chooses which stone \blackstone[1] he wishes to keep on the board and removes the other stone. In this case, the white player has no choice and must keep the stone touching \blackstone[1]. If black had chosen the other stone then white could have chosen either stone \whitestone[2]. The result, shown in Figure 2, is that both players have a single stone left and have removed the stones at \textbf{a} from the board.  

\cleargoban
\black[1]{d3} \white[2]{d2} 
\gobansymbol{c5,e6}{a}
\begin{center}
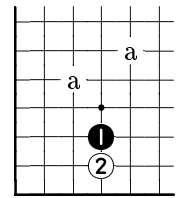
\shortstack{\showgoban\\Figure 2: After the collapse of two pairs of entangled stones}
\end{center}

More generally, when a player (say black) plays an entangled pair of stones on the board, the following collapse process takes place if and only if one of the two stones that have just been placed on the board is touching any other stone. \\
(1) The player who has just played (black) removes one stone from each entangled pair of their colour (not including the pair they last played) which has one or more stones directly adjacent to a stone which has just been placed. \\
(2) The other player (white) then removes one stone from each entangled pair of their colour which has one or more stones directly adjacent to a stone which has just been placed. \\
(3) The player who has just played (black) then removes one of the two entangled stones they have just placed on the board. They must keep a stone that is touching another stone that has just collapsed if they can do so. \\

If at least one stone in an entangled pair is played in contact with a stone which is no longer entangled, then the new pair collapses immediately and play mimics regular Go. A player can always force the collapse of an opponent's pair at the cost of making a potentially undesirable contact play. Similarly, a player can choose to always collapse their own entangled pairs but this will lead to a disadvantage. As the number of stones increases, all the entangled pairs of stones eventually reduce to single stones and Quantum Go increasingly resembles normal Go. \\

The following section will introduce the rules of Quantum Go in more detail. Subsequent sections will illustrate the rules of the game by presenting two example games of Quantum Go, a short game on a small 6 $\times$ 6 board and the first 100 moves of a game on the full 19 $\times$ 19 board. We shall then discuss how Quantum Go provides an interesting background to illustrate some fundamental ideas from quantum theory. Finally, we will conclude by introducing several possible variants of Quantum Go.    

\section{The rules of Quantum Go} 

\textbf{Quantum Go} is played between two players, by placing stones of two different colours-- usually black for one player and white for the other-- on the intersections of a 19 $\times$ 19 checkered square grid. Unless players agree to place a handicap, the board is empty at the onset of the game. Black makes the first move, after which White and Black alternate. A player may pass their turn at any time. \\

A move consists of placing two stones (called \textbf{entangled stones}) of one's own color on two distinct empty intersection on the board. 

When a player plays an entangled pair of stones on the board, the following process (called the \textbf{collapse} of an entangled pair) takes place if and only if there are one or more stones in the intersections neighboring the two stones that have just been placed: \\
(1) The player who has just played removes one stone from each entangled pair of their colour which is already on the board  (not including the pair they last played) and has a stone adjacent to one of the stones which has just been placed. \\
(2) The other player then removes one stone from each entangled pair of their colour which has one or more stones directly adjacent to a stone which has just been placed. \\
(3) The player who has played the last pair of stones then removes one of the two entangled stones they have just placed. They must choose to keep a stone adjacent to a stone that has just collapsed in step (1) or (2) if they can do so. \\

A stone or solidly connected group of stones of one color is captured and removed from the board when all the intersections directly adjacent to it are occupied by stones of the other colour (after a collapse process). No stone may be played so as to recreate a former board position.

Two consecutive passes or a resignation end the game. A player's \textbf{territory} consists of all the points the player has either occupied or surrounded and the player with more territory wins (a komi can also be added to white's total territory). Note that this version resembles the Chinese area counting rules of Go but we could also restate the rules of Quantum Go so that they resemble the Japanese/Korean territory counting rules of Go.

\section{Example 1: a reduced game of Quantum Go}

We will now present a short game of Quantum Go, played on a 6 $\times$ 6 board (with no komi).

In the first two moves, shown in Figure 3, black and white both play entangled pairs. With the third move, black causes the collapse of all three moves by playing a stone adjacent to both $\blackstone[1]$ and \whitestone[2]. The collapse process follows the usual three steps: \\
(1) Black first chooses D3 and takes C2 off the board.  \\
(2) White chooses C4 and takes D5 off the board. \\
(3) Black can choose either stone as they are both adjacent to collapsed stones. He chooses D4 and removes C3 from the board. \\

As a side note, if a game of Quantum Go is played on a regular Go board then both players could place their entangled stones upside down and make a note of the entangled pairs. Once the pair collapses to a single stone, that stone may be flipped so that it is in the normal orientation. \\

\cleargoban
\gobansize{6}
\black[1]{d3} \black[1]{c2}  
\white[2]{c4}   \white[2]{d5} 
\black[3]{d4} \black[3]{c3} 
\begin{center}
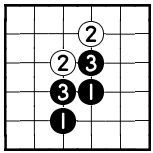
\shortstack{\showgoban\\Figure 3: Start of example game 1}
\end{center}

The game then proceeds almost exactly like a normal game of Go, until move 7, when a new entangled pair is formed, as shown in Figure 4.

\cleargoban
\gobansize{6}
\black[1]{d3,c4,d4,c3,d2,c5,b2} 
\black[7]{e5}  
\begin{center}
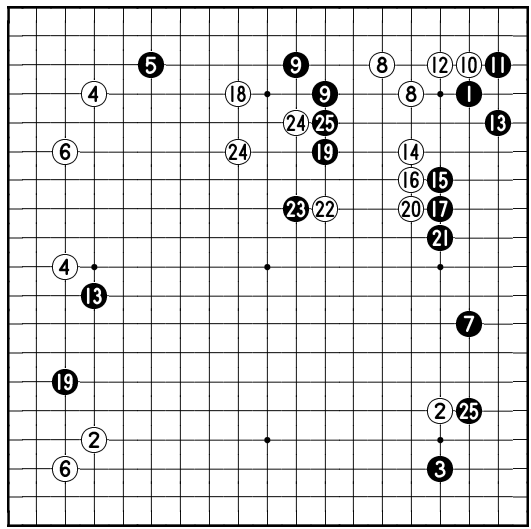
\shortstack{\showgoban\\Figure 4: A new entangled pair is formed}
\end{center}

This entangled pair collapses on the next move since white's entangled pair \whitestone[8] at $\{D5, A3\}$ has a stone in contact with the \blackstone[7] stone at E5. White has no entangled stones in contact with either stone in the entangled pair $\{D5, A3\}$. Black chooses to remove E5 from the board and keep B2. White then chooses to remove the A3 stone from the board and keep D5. The game then proceeds as in Figure 5 until the formation of the final entangled pair at \blackstone[15].   

\cleargoban
\gobansize{6}
\black[1]{d3,c4,d4,c3,d2,c5,b2,d5,b3,c2,c1,e4,e3,f4,e6}
\black[15]{a5}  
\begin{center}
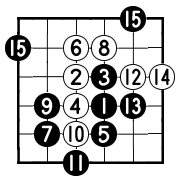
\shortstack{\showgoban\\Figure 5: The final entangled pair}
\end{center}

As before, the pair collapses on the next move. Black chooses to keep E6 and remove A5 from the board. White is then forced to keep E5 -- which is touching the E6 stone that just collapsed -- and must remove B4 from the board. The rest of the game, shown in Figure 6, proceeds exactly like a normal game of Go since all the entangled pairs collapse immediately. \whitestone[22] is at \whitestone[26] and is captured by \blackstone[23]. \whitestone[26] captures three stones.

\cleargoban
\gobansize{6}
\black[1]{d3,c4,d4,c3,d2,c5,b2,d5,b3,c2,c1,e4,e3,f4,e6,e5,b4,b5,a5,b6,d6,f6,f5,c6,f3,f6,a4,a6}  
\begin{center}
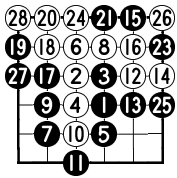
\shortstack{\showgoban\\Figure 6: The game finishes}
\end{center}

Black wins by four points. This example game was a useful simple illustration of the rules but was hardly an interesting game of Quantum Go. Indeed, the small size of the board meant that both players had no choice but to make contact plays from the outset and therefore entangled pairs did not persist throughout the game.

We end this section by  presenting a full record of all the entangled pairs of stones played by both players throughout the game. The move written in bold is the one which made it through to the end of the game and the underlined pairs induced the collapse of another entangled pair (and therefore collapsed immediately).

\begin{center}
Black 1: \textbf{D3} C2 ;
White 2: \textbf{C4} D5 \\
Black 3: \underline{\textbf{D4} C3} ;
White 4: \textbf{C3} D5 \\
Black 5: \textbf{D2} D5 ;
White 6: \textbf{C5} D5 \\
Black 7: \textbf{B2} E5  ;
White 8: \underline{\textbf{D5} A3}  \\
Black 9: \textbf{B3} E5 ;
White 10: \textbf{C2} E4 \\
Black 11: \textbf{C1} E5 ;
White 12: \textbf{E4} E5 \\
Black 13: \textbf{E3} B4 ;
White 14: \textbf{F4} E5 \\
Black 15: \textbf{E6} A5  ;
White 16: \underline{\textbf{E5} B4} \\
Black 17: \textbf{B4} A5 ;
White 18: \textbf{B5} F3 \\
Black 19: \textbf{A5} F3 ;
White 20: \textbf{B6} F3 \\
Black 21: \textbf{D6} F3 ;
White 22: \textbf{F6} B1 \\
Black 23: \textbf{F5} F3 ;
White 24: \textbf{C6} F6 \\
Black 25: \textbf{F3} A4 ;
White 26: \textbf{F6} A4 \\
Black 27: \textbf{A4} A6 ;
White 28: \textbf{A6} A1 \\
Black 29: PASS ;
White 30: PASS \\
\end{center}

\section{Example 2: a full board game of Quantum Go}

We will now proceed to show the first 100 moves of a more interesting full board game of Quantum Go, played on a 19 $\times$ 19 board (with a komi of 7.5 points).

As shown in Figure G1, the opening of a game of Quantum Go is expected to lead to a large number of entangled pairs of stones, as both players tend to avoid the local disadvantage of an early contact play. On move \whitestone[10], white contacts two pairs of entangled white stones and the first collapse process of the game occurs.

\cleargoban
\gobansize{19}
\black[1]{r16} \black[1]{k10}
\white[2]{d4} \white[2]{q5}
\black[3]{q3} \black[3]{o4}
\white[4]{d16} \white[4]{c10} 
\black[5]{f17} \black[5]{q11}
\white[6]{c14} \white[6]{c3} 
\black[7]{r8} \black[7]{c18}
\white[8]{p16} \white[8]{o17}
\black[9]{m16} \black[9]{l17}
\white[10]{r17} \white[10]{r3}
\begin{center}
\shortstack{\showfullgoban\\Figure G1: The game begins}
\end{center}

Black chooses to keep the \blackstone[1] stone at R16 and the \blackstone[3] stone at Q3 and removes the K10 and O4 stones from the board. White can choose to keep either stone in the \whitestone[10] entangled pair as they are both touching another stone that just collapsed. White keeps the R17 stone and removes the R3 stone. 

The game then continues as shown in Figure G2. The \blackstone[11] and \whitestone[12] stones collapse immediately as they are in contact with the collapsed \whitestone[10] stone. 

\cleargoban
\gobansize{19}
\black[1]{r16} 
\white[2]{d4} \white[2]{q5}
\black[3]{q3} 
\white[4]{d16} \white[4]{c10} 
\black[5]{f17} \black[5]{q11}
\white[6]{c14} \white[6]{c3} 
\black[7]{r8} \black[7]{c18}
\white[8]{p16} \white[8]{o17}
\black[9]{m16} \black[9]{l17}
\white[10]{r17} 
\black[11]{s17}
\white[12]{q17}
\black[13]{s15} \black[13]{d9}
\white[14]{p14} \white[14]{j16}
\black[15]{q13} \black[15]{o15}
\white[16]{p13}  \white[16]{d18}
\begin{center}
\shortstack{\showfullgoban\\Figure G2: White pushes from behind}
\end{center}

The \whitestone[16] pair is in contact with three different stones so the collapse process occurs in the usual order: \\
(1) White's entangled pair \whitestone[14] collapses and white chooses to keep the P14 stone. \\  
(2) Black's entangled pairs \blackstone[7] and \blackstone[15] collapse and black chooses to keep the R8 and Q13 stones respectively. \\  
(3) The white entangled pair \whitestone[16] that has just been played collapses and white must keep the P13 stone (as the P13 stone is in contact with another collapsed stone but the D18 stone is not). \\

The game then continues as shown in Figure G3. \blackstone[17] causes the collapse of the \blackstone[5] entangled pair and black chooses to keep F17 on the board.

\cleargoban
\gobansize{19}
\black[1]{r16} 
\white[2]{d4} \white[2]{q5}
\black[3]{q3} 
\white[4]{d16} \white[4]{c10} 
\black[5]{f17} 
\white[6]{c14} \white[6]{c3} 
\black[7]{r8} 
\white[8]{p16} \white[8]{o17}
\black[9]{m16} \black[9]{l17}
\white[10]{r17} 
\black[11]{s17}
\white[12]{q17}
\black[13]{s15} \black[13]{d9}
\white[14]{p14} 
\black[15]{q13}
\white[16]{p13}  
\black[17]{q12}
\white[18]{j16}  \white[18]{h17} 
\black[19]{m14}  \black[19]{c6} 
\white[20]{p12} 
\black[21]{q11} 
\white[22]{m12} \white[22]{o4}
\black[23]{l12} \black[23]{j17} 
\begin{center}
\shortstack{\showfullgoban\\Figure G3: The first fight}
\end{center}

Black \blackstone[23] induces the collapse of the entangled pairs \whitestone[18] and \whitestone[22]. Therefore, white chooses to keep the stones at J16 and M12 and then black chooses the \blackstone[23] stone at L12. \\

Given that fighting usually involves many contact plays, it leads to the frequent collapse of entangled pairs of stones and resembles close combat in normal Go. 
Figures G4 and G5 show that during fighting, collapse is more frequent.

\cleargoban
\gobansize{19}
\black[1]{r16} 
\white[2]{d4} \white[2]{q5}
\black[3]{q3} 
\white[4]{d16} \white[4]{c10} 
\black[5]{f17} 
\white[6]{c14} \white[6]{c3} 
\black[7]{r8} 
\white[8]{p16} \white[8]{o17}
\black[9]{m16} \black[9]{l17}
\white[10]{r17} 
\black[11]{s17}
\white[12]{q17}
\black[13]{s15} \black[13]{d9}
\white[14]{p14} 
\black[15]{q13}
\white[16]{p13}  
\black[17]{q12}
\white[18]{j16}   
\black[19]{m14}  \black[19]{c6} 
\white[20]{p12} 
\black[21]{q11} 
\white[22]{m12} 
\black[23]{l12} 
\white[24]{l15} \white[24]{j14}
\black[25]{m15} \black[25]{r5} 
\begin{center}
\shortstack{\showfullgoban\\Figure G4: Fighting continues}
\end{center}

Black \blackstone[25] first induces the collapse of \blackstone[9] and \blackstone[19], where black chooses M16 and M14, and then causes the collapse of \whitestone[2] and \whitestone[24], where white chooses D4 and M15. This means that black must keep the stone at L15, which is in contact with the collapsed stone \whitestone[24]. \\

In Figure G5, we only show the stones that both players chose to keep when their entangled pairs collapsed. Note that black removed the \blackstone[29] and \blackstone[31] stones at \blackstone[31] and \blackstone[33] respectively. Similarly, white removed the  \whitestone[30] and \whitestone[32] stones  at \textbf{w}.

Also, \whitestone[30] induced the collapse of \whitestone[8] and white chose to keep the \whitestone[8] stone at P16.

\cleargoban
\gobansize{19}
\black[1]{r16} 
\white[2]{d4} 
\black[3]{q3} 
\white[4]{d16} \white[4]{c10} 
\black[5]{f17} 
\white[6]{c14} \white[6]{c3} 
\black[7]{r8} 
\white[8]{p16} 
\black[9]{m16} 
\white[10]{r17} 
\black[11]{s17}
\white[12]{q17}
\black[13]{s15} \black[13]{d9}
\white[14]{p14} 
\black[15]{q13}
\white[16]{p13}  
\black[17]{q12}
\white[18]{j16}   
\black[19]{m14}  
\white[20]{p12} 
\black[21]{q11} 
\white[22]{m12} 
\black[23]{l12} 
\white[24]{l15}
\black[25]{m15} 
\white[26]{m13}
\black[27]{l13} 
\white[28]{l14}
\black[29]{o15,p15,k17,j17,k16,j15,k15,k14,j14,k13,m11,n11,h14}
\white[42]{j12} \white[42]{j10}
\black[43]{n18} \black[43]{f15}
\gobansymbol{l11}{w}
\begin{center}
\shortstack{\showfullgoban\\Figure G5: A group between life and death}
\end{center}

The move \blackstone[43] leads to an interesting situation where the black group on the top side is locally in a \textbf{superposition} of life and death depending on how black will later choose to collapse the entangled pair \blackstone[43]. \\

The game proceeds with an interesting fight, shown in Figure G6, where black aims to counter-attack by threatening the white group on the right. From this point onwards, we will only show a single stone when an entangled pair collapses as soon as it is placed on the board and does not induce the collapse of another entangled pair.

\cleargoban
\gobansize{19}
\black[1]{r16} 
\white[2]{d4} 
\black[3]{q3} 
\white[4]{d16} \white[4]{c10} 
\black[5]{f17} 
\white[6]{c14} \white[6]{c3} 
\black[7]{r8} 
\white[8]{p16} 
\black[9]{m16} 
\white[10]{r17} 
\black[11]{s17}
\white[12]{q17}
\black[13]{s15} \black[13]{d9}
\white[14]{p14} 
\black[15]{q13}
\white[16]{p13}  
\black[17]{q12}
\white[18]{j16}   
\black[19]{m14}  
\white[20]{p12} 
\black[21]{q11} 
\white[22]{m12} 
\black[23]{l12} 
\white[24]{l15}
\black[25]{m15} 
\white[26]{m13}
\black[27]{l13} 
\white[28]{l14}
\black[29]{o15,p15,k17,j17,k16,j15,k15,k14,j14,k13,m11,n11,h14}
\white[42]{j12} \white[42]{j10}
\black[43]{n18} \black[43]{f15}
\white[44]{g16} \white[44]{q7}
\black[45]{l11} 
\white[46]{n10}
\black[47]{l9} \black[47]{l4}
\white[48]{j18} 
\black[49]{p18} \black[49]{l6}
\white[50]{n8} \white[50]{p9}
\black[51]{m7} \black[51]{q8}
\begin{center}
\shortstack{\showfullgoban\\Figure G6: Black counter-attacks}
\end{center}

Black \blackstone[51] induces the collapse of \whitestone[44]. White chooses to keep the \whitestone[44] stone at G16 and then black chooses to keep the \blackstone[51] stone at M7.

Figure G7 shows how white forces black to collapse his entangled pairs in order to make his group live locally. 

\cleargoban
\gobansize{19}
\black[1]{r16} 
\white[2]{d4} 
\black[3]{q3} 
\white[4]{d16} \white[4]{c10} 
\black[5]{f17} 
\white[6]{c14} \white[6]{c3} 
\black[7]{r8} 
\white[8]{p16} 
\black[9]{m16} 
\white[10]{r17} 
\black[11]{s17}
\white[12]{q17}
\black[13]{s15} \black[13]{d9}
\white[14]{p14} 
\black[15]{q13}
\white[16]{p13}  
\black[17]{q12}
\white[18]{j16}   
\black[19]{m14}  
\white[20]{p12} 
\black[21]{q11} 
\white[22]{m12} 
\black[23]{l12} 
\white[24]{l15}
\black[25]{m15} 
\white[26]{m13}
\black[27]{l13} 
\white[28]{l14}
\black[29]{o15,p15,k17,j17,k16,j15,k15,k14,j14,k13,m11,n11,h14}
\white[42]{j12} \white[42]{j10}
\black[43]{n18} \black[43]{f15}
\white[44]{g16} \white[44]{q7}
\black[45]{l11} 
\white[46]{n10}
\black[47]{l9} \black[47]{l4}
\white[48]{j18} 
\black[49]{p18} \black[49]{l6}
\white[50]{n8} \white[50]{p9}
\black[51]{m7} 
\white[52]{o17} \white[52]{l18}
\black[53]{o18} \black[53]{d17}
\begin{center}
\shortstack{\showfullgoban\\Figure G7: White forces black to commit}
\end{center}

\whitestone[52] is a sharp forcing move and black \blackstone[53] induces a collapse process. Black first chooses to keep the \blackstone[43] and \blackstone[49] stones at N18 and P18 respectively. White then chooses to keep the \whitestone[4] and \whitestone[44] stones at D16 and O17 respectively. Finally black removes the D17 \blackstone[53] stone from the board. \\

The game then continues as in Figure G8. White \whitestone[56] causes the collapse of \whitestone[50] and white removes the stone at P9. The entangled pair \whitestone[58] at $\{L8, Q7\}$ causes the collapse of the \blackstone[47] pair. Black chooses to remove the \blackstone[47] stone at \textbf{b} and white must remove the stone \whitestone[58] at \textbf{w} from the board.

Black \blackstone[61] induces the collapse of the entangled pair \whitestone[42] and white removes the stone at \textbf{x}. 

\cleargoban
\gobansize{19}
\black[1]{r16} 
\white[2]{d4} 
\black[3]{q3} 
\white[4]{d16}
\black[5]{f17} 
\white[6]{c14} \white[6]{c3} 
\black[7]{r8} 
\white[8]{p16} 
\black[9]{m16} 
\white[10]{r17} 
\black[11]{s17}
\white[12]{q17}
\black[13]{s15} \black[13]{d9}
\white[14]{p14} 
\black[15]{q13}
\white[16]{p13}  
\black[17]{q12}
\white[18]{j16}   
\black[19]{m14}  
\white[20]{p12} 
\black[21]{q11} 
\white[22]{m12} 
\black[23]{l12} 
\white[24]{l15}
\black[25]{m15} 
\white[26]{m13}
\black[27]{l13} 
\white[28]{l14}
\black[29]{o15,p15,k17,j17,k16,j15,k15,k14,j14,k13,m11,n11,h14}
\white[42]{j12} 
\black[43]{n18} 
\white[44]{g16} 
\black[45]{l11} 
\white[46]{n10}
\black[47]{l9} 
\white[48]{j18} 
\black[49]{p18} 
\white[50]{n8} 
\black[51]{m7} 
\white[52]{o17} 
\black[53]{o18} 
\white[54]{l18} \white[54]{c17}
\black[55]{m17}
\white[56]{n7}
\black[57]{m6}
\white[58]{l8}
\black[59]{m8}
\white[60]{m9}
\black[61]{k10} 
\white[62]{k8}
\black[63]{h10} \black[63]{h8}
\white[64]{h11} \white[64]{k9}
\gobansymbol{l4}{b}
\gobansymbol{q7}{w}
\gobansymbol{j10}{x}
\begin{center}
\shortstack{\showfullgoban\\Figure G8: White cuts in the center}
\end{center}

The white pair \whitestone[64] induces the collapse of the \blackstone[63] so black chooses to remove the H8 stone and then white must remove the \whitestone[64] stone at K9 from the board. \\

The game continues as shown in Figure G9 and both players push down in order to settle their groups. White \whitestone[78] causes the collapse of black \blackstone[77] and therefore black removes the F3 stone from the board and then white chooses to remove the F4 stone (which is no longer in contact with a collapsed stone).

\cleargoban
\gobansize{19}
\black[1]{r16} 
\white[2]{d4} 
\black[3]{q3} 
\white[4]{d16}
\black[5]{f17} 
\white[6]{c14} \white[6]{c3} 
\black[7]{r8} 
\white[8]{p16} 
\black[9]{m16} 
\white[10]{r17} 
\black[11]{s17}
\white[12]{q17}
\black[13]{s15} \black[13]{d9}
\white[14]{p14} 
\black[15]{q13}
\white[16]{p13}  
\black[17]{q12}
\white[18]{j16}   
\black[19]{m14}  
\white[20]{p12} 
\black[21]{q11} 
\white[22]{m12} 
\black[23]{l12} 
\white[24]{l15}
\black[25]{m15} 
\white[26]{m13}
\black[27]{l13} 
\white[28]{l14}
\black[29]{o15,p15,k17,j17,k16,j15,k15,k14,j14,k13,m11,n11,h14}
\white[42]{j12} 
\black[43]{n18} 
\white[44]{g16} 
\black[45]{l11} 
\white[46]{n10}
\black[47]{l9} 
\white[48]{j18} 
\black[49]{p18} 
\white[50]{n8} 
\black[51]{m7} 
\white[52]{o17} 
\black[53]{o18} 
\white[54]{l18} \white[54]{c17}
\black[55]{m17}
\white[56]{n7}
\black[57]{m6}
\white[58]{l8}
\black[59]{m8}
\white[60]{m9}
\black[61]{k10} 
\white[62]{k8}
\black[63]{h10} 
\white[64]{h11} 
\black[65]{g10,g14,j11,k12,j10,n6,m5,n5,m4,n4,m3,n3}
\black[77]{j7} \black[77]{f3} 
\white[78]{r5} \white[78]{f4}
\begin{center}
\shortstack{\showfullgoban\\Figure G9: Both players settle their groups}
\end{center}

As shown in Figure G10, black then makes a double approach on white's corner in the bottom left. White \whitestone[82] induces the collapse of the \blackstone[81] pair and black chooses to remove the stone at F15, which forces white to remove the stone at E12. 

\cleargoban
\gobansize{19}
\black[1]{r16} 
\white[2]{d4} 
\black[3]{q3} 
\white[4]{d16}
\black[5]{f17} 
\white[6]{c14} \white[6]{c3} 
\black[7]{r8} 
\white[8]{p16} 
\black[9]{m16} 
\white[10]{r17} 
\black[11]{s17}
\white[12]{q17}
\black[13]{s15} \black[13]{d9}
\white[14]{p14} 
\black[15]{q13}
\white[16]{p13}  
\black[17]{q12}
\white[18]{j16}   
\black[19]{m14}  
\white[20]{p12} 
\black[21]{q11} 
\white[22]{m12} 
\black[23]{l12} 
\white[24]{l15}
\black[25]{m15} 
\white[26]{m13}
\black[27]{l13} 
\white[28]{l14}
\black[29]{o15,p15,k17,j17,k16,j15,k15,k14,j14,k13,m11,n11,h14}
\white[42]{j12} 
\black[43]{n18} 
\white[44]{g16} 
\black[45]{l11} 
\white[46]{n10}
\black[47]{l9} 
\white[48]{j18} 
\black[49]{p18} 
\white[50]{n8} 
\black[51]{m7} 
\white[52]{o17} 
\black[53]{o18} 
\white[54]{l18} \white[54]{c17}
\black[55]{m17}
\white[56]{n7}
\black[57]{m6}
\white[58]{l8}
\black[59]{m8}
\white[60]{m9}
\black[61]{k10} 
\white[62]{k8}
\black[63]{h10} 
\white[64]{h11} 
\black[65]{g10,g14,j11,k12,j10,n6,m5,n5,m4,n4,m3,n3,j7,r5}
\black[79]{f3} \black[79]{e14}
\white[80]{g13}
\black[81]{c6} \black[81]{f15}
\white[82]{d6} \white[82]{e12}
\begin{center}
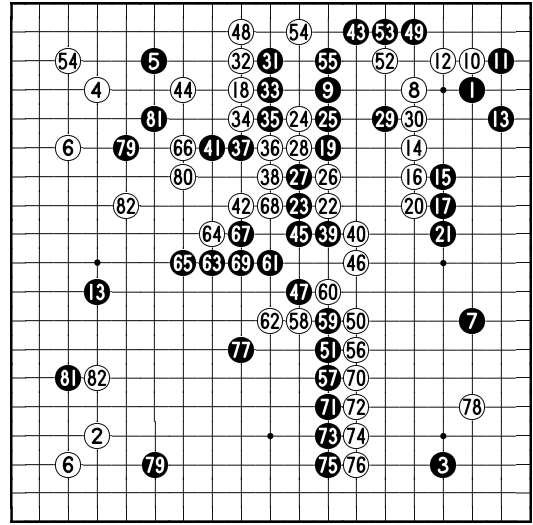
\shortstack{\showfullgoban\\Figure G10: Double approach in the lower left}
\end{center}

White then plays an interesting joseki (corner variation) in the bottom left, which can be seen in Figure G11. White \whitestone[84] induces the collapse of the entangled pair \blackstone[79] and therefore black removes the \blackstone[79] stone at \textbf{b} and white removes the \whitestone[84] stone at \textbf{w}. Black \blackstone[93] induces the collapse of the entangled pair \blackstone[13] and Black removes the \blackstone[13] stone at \textbf{x}. The final collapse process of the 100 moves we will show occurs when \blackstone[97] touches the white stone \whitestone[54]. White chooses to remove the \whitestone[54] stone at \textbf{y} and black must keep the stone at K18.

\cleargoban
\gobansize{19}
\black[1]{r16} 
\white[2]{d4} 
\black[3]{q3} 
\white[4]{d16}
\black[5]{f17} 
\white[6]{c14} \white[6]{c3} 
\black[7]{r8} 
\white[8]{p16} 
\black[9]{m16} 
\white[10]{r17} 
\black[11]{s17}
\white[12]{q17}
\black[13]{s15} 
\white[14]{p14} 
\black[15]{q13}
\white[16]{p13}  
\black[17]{q12}
\white[18]{j16}   
\black[19]{m14}  
\white[20]{p12} 
\black[21]{q11} 
\white[22]{m12} 
\black[23]{l12} 
\white[24]{l15}
\black[25]{m15} 
\white[26]{m13}
\black[27]{l13} 
\white[28]{l14}
\black[29]{o15,p15,k17,j17,k16,j15,k15,k14,j14,k13,m11,n11,h14}
\white[42]{j12} 
\black[43]{n18} 
\white[44]{g16} 
\black[45]{l11} 
\white[46]{n10}
\black[47]{l9} 
\white[48]{j18} 
\black[49]{p18} 
\white[50]{n8} 
\black[51]{m7} 
\white[52]{o17} 
\black[53]{o18} 
\white[54]{l18} 
\black[55]{m17}
\white[56]{n7}
\black[57]{m6}
\white[58]{l8}
\black[59]{m8}
\white[60]{m9}
\black[61]{k10} 
\white[62]{k8}
\black[63]{h10} 
\white[64]{h11} 
\black[65]{g10,g14,j11,k12,j10,n6,m5,n5,m4,n4,m3,n3,j7,r5}
\black[79]{f3} 
\white[80]{g13}
\black[81]{c6} 
\white[82]{d6} 
\black[83]{d7}
\white[84]{f4} 
\black[85]{g4,g3,f2,c5,e6,d5,e4,c7,d8,m2,l2,s18,k18,k19,n14,n12}
\gobansymbol{e14}{b}
\gobansymbol{r10}{w}
\gobansymbol{d9}{x}
\gobansymbol{c17}{y}
\begin{center}
\shortstack{\showfullgoban\\Figure G11: One hundred moves}
\end{center}

We expect that these one hundred moves are a sufficient illustration of the rules of Quantum Go. 

\newpage
\section{Quantum metaphors}

Quantum Go has been designed to closely resemble regular Go, particularly in close combat situations and in the later stages of the game. Quantum Go, however, adds a level of complexity by allowing players to explore the possibilities coming from stones being in two places at once. This phenomenon simulates \textbf{quantum superposition}, which is the principle that physical objects are not required to occupy a precise position. Moreover, even entire groups of stones in Quantum Go can be neither alive nor dead, but in a superposition of both states.

In quantum theory, physical systems exhibiting superposition, such as quantum `particles' in the two-slit experiment \cite{Fey62}, are studied through a quantum measurement. This can be interpreted as a phenomenon of \textbf{quantum collapse}, where the physical states of a system which are in a superposition can be reduced to states which are no longer in a superposition. In Quantum Go, an analogous process occurs when a Go stone is in contact with one of the stones in an entangled pair. When the contact takes place, the pair of go stones is no longer in a quantum-like superposition but undergoes a process akin to quantum collapse, which leads to each stone in the pair either being present or absent on the board.

Another quantum concept that can be illustrated through Quantum Go is \textbf{entanglement}. Quantum entanglement is a phenomenon where pairs or groups of physical systems are generated or interact in ways where the quantum state of each subsystem cannot be described independently. Instead, an entangled quantum state must be described for the system as a whole. As the name indicates, the stones in a pair of entangled go stones cannot be described independently (even though they are dispersed in space) until after they have undergone the collapse process. \\

We can use mathematics to describe these ideas more precisely. Let us consider an intersection \textbf{x} on the square grid of a Go board. In regular Go, each intersection \textbf{x} can either have a black stone, a white stone or be empty. This means that the a subsystem of the whole board corresponding to the intersection \textbf{x} is either in the state $(\ket{1}_{x})_B$, when there is a black stone at \textbf{x}, in the state $(\ket{1}_{x})_W$, when there is a white stone at \textbf{x}, or in the state $\ket{0}_{x}$, when there is no stone at \textbf{x}.

We can also look at pairs of intersections \textbf{x} and \textbf{y} and express the subsystem of the whole board corresponding to these two intersections. In regular Go, there are nine possible states for these intersections which are the following: \\
$\ket{0}_{x} \ket{0}_{y}$, where there is no stone placed at either intersection. \\
$\ket{0}_{x} (\ket{1}_{y})_B$, where there is a black stone at \textbf{y} and no stone at \textbf{x}. \\
$\ket{0}_{x} (\ket{1}_{y})_W$, where there is a white stone at \textbf{y} and no stone at \textbf{x}. \\
$(\ket{1}_{x})_B \ket{0}_{y}$, where there is a black stone at \textbf{x} and no stone at \textbf{y}. \\
$(\ket{1}_{x})_W \ket{0}_{y}$, where there is a white stone at \textbf{x} and no stone at \textbf{y}. \\
 $(\ket{1}_{x})_B (\ket{1}_{y})_B$, $(\ket{1}_{x})_W (\ket{1}_{y})_W$, $(\ket{1}_{x})_B (\ket{1}_{y})_W$ and $(\ket{1}_{x})_W (\ket{1}_{y})_B$, where there are stones at both intersections. \\  

In Quantum Go, however, a player no longer plays single go stones but places pairs of entangled go stones instead. Therefore, the nine states above only arise for pairs of board intersections containing stones that have already collapsed. When an entangled pair of black stones is placed on a pair of intersections \textbf{x} and \textbf{y}, then the intersections are in the state:\\
\begin{center}
$(\alpha \ket{0}_{x} \ket{1}_{y} - \beta \ket{1}_{x} \ket{0}_{y})_B$\\
\end{center}
 where the (complex) coefficient $\alpha$ indicates the likelihood that there will be a stone at \textbf{y} but not at \textbf{x} following the collapse process and the (complex) coefficient $\beta$ indicates the likelihood that there will be a stone at \textbf{x} but not at \textbf{y}. This resembles the behavior of a pair of entangled quantum systems and explains our use of the term \textbf{entangled pair} of stones.

The subsystem of the whole board corresponding to a single intersection \textbf{x}, where there is one black stone in an entangled pair, can be written as a (mixed) quantum state: \\ 
\begin{center}
$\vert \alpha \vert^2 \ket{0}_x \bra{0}_x + \vert \beta \vert^2 (\ket{1}_x \bra{1}_x)_B$ \\
\end{center}
where $\vert \alpha \vert^2$ and $\vert \beta \vert^2$ are (real) coefficients which describe the likelihood that intersection \textbf{x} will be chosen after collapse instead of the intersection with the other entangled stone. If we have no information, then $\vert \alpha \vert^2 = \vert \beta \vert^2 =\frac{1}{2}$ and it is equally likely that there will or will not be a stone at intersection \textbf{x} following the collapse process. 

This shows that in Quantum Go, the state of each intersection is in a quantum superposition.  

We can then write the state of the whole board as the combination (tensor product) of the states of all the intersections containing empty spaces, single stones and entangled pairs. For example, the tiny board shown in Figure 7 can be given the state: \\
\scriptsize
\begin{center}
$ (\ket{0}_{A1,A3,B2,B3,C1,C3}) (\ket{1}_{C2})_{W} (\alpha_1 \ket{0}_{A2} \ket{1}_{B1} - \beta_1 \ket{1}_{A2} \ket{0}_{B1})_{B}  $\\
\end{center}
\normalsize

\cleargoban
\gobansize{3}
\black[1]{a2} \black[1]{b1}  
\white[2]{c2}   
 
\begin{center}
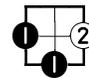
\shortstack{\showgoban\\Figure 7: State of a 3 $\times$ 3 board}
\end{center}

This provides an intuitive illustration of quantum states and how subsystems can combine to create larger systems.   

The collapse of an entangled pair also mimics a quantum measurement. Given the state $(\alpha \ket{0}_{x} \ket{1}_{y} - \beta \ket{1}_{x} \ket{0}_{y})_W$ for a pair of intersections on the board which contain a white entangled pair, the black player does not know which stone will be taken off the board. Therefore, from his perspective, the black player is performing a type of (projective) quantum measurement by touching the opponent's stone with his own. By placing a stone adjacent to the entangled stone at intersection \textbf{x}, the black player is performing a measurement  $\{\ket{0}_{x} \bra{0}_x , (\ket{1}_x \bra{1}_x)_W \}$, where he ignores which outcome will occur.

We should mention that the most interesting quantum phenomena, including complementarity \cite{Bo35}, non-locality \cite{Bru14} and contextuality \cite{Ko67} have no analogues in Quantum Go. Correlations between measurements on subsystems are indistinguishable from any other classical correlation between classical random variables and obey the Bell and CHSH inequalities \cite{Cla69}. In fact Quantum Go only mimics the probabilistic behavior of quantum theory due to each player's ignorance concerning the opponent's decisions. Nevertheless, playing Quantum Go provides a valuable insight into some of the fundamental phenomena of quantum mechanics. \\

\section{Conclusion}

In this article, we have introduced the rules of Quantum Go and illustrated these rules by presenting two example games. In addition to being a fun and interesting game in its own right, Quantum Go has been designed to make the nature of quantum mechanics more accessible to everyone. It is well suited for scientific outreach activities or for a classroom environment.

We can imagine many variations of Quantum Go which simplify the game and make it easier to play. A natural modification is \textbf{Weak Quantum Go}, where the collapse process occurs whenever a stone is placed in an intersection either adjacent or directly diagonal from one or more other stones. Another interesting variant is \textbf{Symmetric Quantum Go}, where entangled stones must be placed at pairs of points which are point-symmetric with respect to the central K10 point (such as C6 and R14). If one of the two opposite points where a player wishes to play are occupied, then the player is forced to place a single stone. Finally, one can imagine \textbf{Semi-Quantum Go}, where black is permitted to play entangled pairs of stones but white is only allowed to place single stones, as in regular Go. It would then be interesting to decide the expected advantage, in terms of points, conferred to the player who can play entangled pairs of stones and adjust the komi accordingly.

Another possibility is to devise variants of the game which are theoretically interesting but might be more complex to play. For example, we can imagine a version which involves allowing entangled groups of stones which are larger than two. The entangled pairs we introduced resemble quantum singlet states \cite{Be04}. If we allow \textbf{entangled triplets} of stones then we could introduce two types of entangled triplets corresponding to GHZ-type and W-type entangled states \cite{Dur00}. The former would then directly collapse to a single stone following the collapse process induced by a touching stone, whereas the later would only collapse into an entangled pair which would later collapse into a single stone. The game could also be generalized to include entangled groups of stones containing more members.

Finally, it would be interesting to create a version of Quantum Go where the stones can be entangled in several complementary ways. It might be possible to have stones of many different colours or with arrows so that they can be placed on the board in different orientations. These variants, although they would no longer resemble the game of Go, could provide interesting toy models to study quantum complementarity.

\bibliographystyle{plainnat}
\bibliography{gobib}

\end{document}